\documentclass[amssymb,floatfix,10pt,twocolumn,
superscriptaddress,nofootinbib]{revtex4}
\usepackage[utf8]{inputenc}
\usepackage{latexsym,float}
\usepackage{epsfig}
\usepackage{epstopdf}
\usepackage{amsmath,hyperref}
\usepackage{amssymb}
\usepackage{graphicx}
\usepackage[T1]{fontenc}

\begin{document}
\title{The effect of vacuum polarization on the magnetic fields \\ around a Schwarzschild black hole }
\author{Petar Pavlović}\email{petar.pavlovic@desy.de}
\affiliation{Institut f\"{u}r Theoretische Physik, Universit\"{a}t
Hamburg, Luruper Chaussee 149, 22761 Hamburg, Germany}
\author{Marko Sossich}
\email{marko.sossich@fer.hr}
\affiliation{University of Zagreb, Faculty of Electrical Engineering and Computing, Department of Physics,
	Unska 3, 10 000 Zagreb, Croatia}

\begin{abstract}
  {\bf Abstract:}  It is a well known result that the 
effect of vacuum polarization in gravitational fields
will lead to a non-minimal coupling between gravity
and electromagnetism. We investigate this phenomenon 
further by considering the description of static
magnetic field around a Schwarzschild black hole.
It is found that close to the Schwarzschild horizon 
the magnetic fields can be strongly modified 
with respect to both cases of magnetic fields on 
flat spacetime
and magnetic fields minimally coupled on curved
spacetime. Under the proper sign of the non-minimal 
coupling parameter, $q$, the effective fields can 
undergo large amplifications. Furthermore, we 
discuss the physical meaning of the singularities 
that arise in the considered problem. We conclude
by discussing the potential observational effects
of vacuum polarization on the magnetic fields.
In the case of astrophysical black holes, depending on the value of the coupling parameter, significant 
modifications of the magnetic near the black hole 
horizons are possible -- which could be used to detect
the vacuum polarization effect or at least to put
constraints on the values of the coupling parameter.
Moreover, we show how the considered effect directly constraints the viability of primordial black holes of sizes smaller than that of the Compton wavelength for the electron, and also impacts the distribution of
magnetic fields in the early Universe.
\end{abstract}
\maketitle

\section{Introduction}

Although the description of electrodynamic 
phenomena in
  terms of Maxwell's equations represents one of the most
  successful physical
theories, it is known that for high enough electromagnetic fields, quantum radiative corrections should become important \cite{weinberg}.
Vacuum polarization is an effect of appearance of virtual pairs in a ``fermion loop'', which changes the propagation properties of
photons in vacuum. This process can be understood as a result of quantum fluctuations, in which a photon can fluctuate into a fermion pair,
described by some probability amplitude which will in general be different from zero. These effects are theoretically well founded
and known for a long time, but their observation still remains to be an open experimental goal \cite{heinzl}. Moreover, the observational
aspects of vacuum polarization were mostly focused on the Minkowski (flat) spacetime. \\ \\
While studying the electromagnetic fields on curved spacetimes in the context of general relativity \cite{tsagas, kand}, 
the focus is mostly put on the macroscopic and classical aspects of these fields. In this framework, the electromagnetic field
is represented by the electromagnetic tensor, which contributes to the stress-energy tensor, thus leading to spacetime curvature, in the
same fashion as other forms of energy distributions. The equations of motion are derived from the Lagrangian density consisting of Einstein-Hilbert and Maxwell contribution,
and the coupling between the electrodynamics and gravitational sector is therefore only minimal. \\ \\
However, for strong enough electromagnetic and gravitational fields the analysis of electromagnetism on curved spacetimes should be modified
to take into account the effects coming from the vacuum polarization. The dominant process will be given by the transition of a photon 
into a $e^{+} e^{-}$ pair. This fluctuation defines a characteristic length scale for the photon, which is of the order of the Compton
wavelength of the electron. Since it now has an effective characteristic length, the photon will be influenced by the change in geometry along this characteristic
scale. This means that properties of photon in vacuum will be influenced by the spacetime curvature, coming from tidal effects on its
length, which are effectively acquired through the vacuum polarization. This phenomenon was studied in \cite{drummond} in the one-loop approach, where it was shown to introduce the
non-minimal coupling between the electromagnetic field and curvature. Apart from the standard Maxwell part, the Lagrangian for electrodynamics
now also consists of the additional contribution, given by the contraction of electromagnetic and curvature tensors, leading to gauge and curvature
invariant terms. This effect also has some important implications on the regime of validity of General relativity. One of the central
principles of General relativity is the Equivalence principle, which is assuming the equivalence of gravitational and inertial effects for a 
sufficiently small regions of spacetime, such that tidal effects can be ignored. However, it is obvious that tidal effects cannot be ignored
even on the scale of electron Compton wavelength, when the vacuum polarization becomes important. \\ \\
In \cite{drummond} the QED vacuum polarization on the curved spacetime was discussed for the simplest case of a photon propagating in vacuum. 
It is not simple to generalize these results to some arbitrary field configurations, but it seems very plausible that the same type
of coupling between electromagnetic and curvature terms will remain to be valid in general, at least in the leading order. However, the coupling
coefficients should be considered as different, and they will probably depend on the characteristic physical regime. In principle, for strong electromagnetic and gravitational fields, characterized by a complex configuration of photons, the characteristic coupling between the electromagnetic and gravitational field could then be much stronger than the one-photon correction, which is of the order of the square of the Compton wavelength for the electron. 
Non-minimal coupling between electrodynamics and gravitational sector was in the recent decades investigated in various theoretical settings \cite{turner, lemos, bamba, prasanna1, prasanna2, sert, sert2, sert3, sert4}.  \\ \\ 
Although the effects of vacuum polarization on curved spacetime electrodynamics are of fundamental significance for our understanding
of physical interactions, and are the first step towards the unification of electromagnetism and gravity, further research on this important
issue is limited by the fact that these effects are quantitatively negligible in most of the observational settings. They could, however,
become important in the high curvature regimes where electromagnetic fields are present, for instance in the very early universe and around
black holes. Due to the very high conductivity of the Universe, electric fields can be taken as vanishing, while
in the same time magnetic fields are known to be present on all scales of the observable Universe -- from planets, to galaxy clusters, and
voids of intergalactic medium \cite{beck,bernet}. It therefore seems necessary that any empirical confirmation of vacuum polarization leading
to a non-minimal coupling between electrodynamics and gravity will be based on the observations of electromagnetic fields in these settings. Probably the
most promising strategy for the potential detection of these effects is the analysis of deformation of galactic magnetic field near the black hole horizons, 
which is in principle observable, for instance using the Zeeman splitting method. Conversely, the absence of such effects in
measurements could be used to set constraints on the values of coupling coefficients between electromagnetism and gravity. The vacuum polarization
could become especially important for primordial black holes, thus changing the evolution of magnetic fields in the early universe and its
large scale distribution. In this work we propose these directions for further study of non-minimal interaction between electromagnetism
and gravity, by analyzing the problem of modification of galactic field around a black hole that comes as its consequence. 
This paper is organized as follows: in section II. we
briefly present the theory of non-minimal coupling 
of gravity and electromagnetism. In section III. the 
relevant equations of motion are derived with the
corresponding assumptions. In section
IV. the equations of motion are solved and its physical
content is analyzed. In section V. some applications on
realistic physical systems are presented and in section
VI. we conclude our work.

\section{Non-minimal coupling of gravity and electromagnetism}
The effects coming from the virtual photon loops, leading to the vacuum polarization in the presence of gravitational and electromagnetic field, can be described by adding a corresponding quantum correction term, $S_{corr}$, to the electromagnetic, $S_{em}$, and gravitational action, $S_{grav}$, so that the total action is given by $S=S_{em}+ S_{grav}$ + $S_{corr}$, where \cite{drummond}
\begin{equation}
S_{corr}= \sum_{n, even}\frac{1}{n!} \int {\displaystyle \prod_{i=1}^{n}} d^{4} x_{i} A_{\mu_{i}}(x_{i}) G^{\mu_{1}...\mu_{n}}(x_{1}...x_{n}) .
\end{equation}
Here $A_{\mu}$ is the gauge $U(1)$ potential and $G^{\mu_{1}...\mu_{n}}$ 
represents the sum over one-particle-irreducible Feynmann diagrams. When this correction is computed to the lowest order in powers of the Compton wavelength, and only the quadratic terms in $A_{\mu}$ considered, the  
total action can be written as \cite{drummond, balakin}
\begin{equation}
 S=\int d^{4}x \sqrt{-g} \mathcal{L},
\end{equation}
where the Lagrangian density is
\begin{equation}
 \mathcal{L}=\frac{R}{\kappa} + \frac{1}{2}F^{\mu\nu}F_{\mu \nu} + \frac{1}{2}\mathcal{R}^{\mu\nu\rho\sigma} F_{\mu \nu}F_{\rho \sigma}+
 \mathcal{L}_{matter},
\end{equation}
where $\kappa=8 \pi G/c^{4}$ (from now on we will set $c=G=1$), $g$ is the determinant of the metric tensor, $R$ is the Ricci scalar, $F^{\mu \nu}$ is the Maxwell tensor obeying $F^{\mu \nu}=\nabla^{\mu}A^{\nu} - \nabla^{\nu}A^{\mu}$ where $\nabla_{\mu}$ is the covariant derivative and $\mathcal{L}_{matter}$ is the Lagrangian of neutral matter. The impact of the vacuum polarization is given through the tensor defined as
\begin{multline}
 \mathcal{R}^{\mu\nu\rho\sigma}\equiv \frac{q_{1}}{2}(g^{\mu\rho}g^{\nu \sigma} - g^{\mu\sigma}g^{\nu\rho})R \\ \\ + \frac{q_{2}}{2}(R^{\nu\rho} g^{\nu\sigma} - R^{\mu\sigma} g^{\nu\rho} + R^{\nu \sigma}g^{\mu\rho} - R^{\nu\rho}g^{\mu\sigma})  +
 q_{3}R^{\mu\nu\rho\sigma}, 
 \label{oneloop}
\end{multline}
where $q_{1}$, $q_{2}$ and $q_{3}$ are the the coupling constants, and as usual $R^{\mu \nu }$ is the Ricci tensor and $R^{\mu\nu\rho\sigma}$ is the Riemann tensor. We can thus clearly see that the considered correction will lead to the non-minimal coupling between electromagnetism and gravity. For the simplest case of a single photon propagation, the coupling constants will naturally be of the order of the Compton wavelength and their values were computed in \cite{drummond}. It is natural to generalize this type of quantum correction, which leads to the non-minimal coupling between gravitational and electromagnetic tensors, to the more general and complex field configurations. 
Thus we will assume that the same type of coupling will remain valid in arbitrary settings, at least to the leading order, but with different values of the coupling coefficients. 
Now it is straightforward to obtain the equations of motion in the non-minimal coupled gravity and electromagnetism, as they are given by varying the action with respect to the $U(1)$ gauge potential $A_{\mu}$. By doing so and by rewriting the equations in the familiar form we get
\begin{equation}
 \nabla_{\mu} (F^{\mu\nu} + \mathcal{R}^{\mu\nu\rho\sigma}F_{\rho\sigma}) =0.
 \label{sacuvanje}
\end{equation}
By specifying the metric tensor $g^{\mu\nu}$ and the Maxwell tensor $F^{\mu\nu}$ one can solve the equation \eqref{sacuvanje} and inspect the behavior of the solution in
the non-minimal coupled theory of gravity and electromagnetism.

\section{Equations of motion in the Schwarzschild metric}

As discussed before, we are interested in the influence of the static spherically symmetric spacetime on the electromagnetic field. Since the astrophysical electric fields can be taken as vanishing, as discussed in the introduction, we set the electric component of the Maxwell tensor to zero and the remaining fields are only the magnetic fields. We will also assume that the magnetic fields are weak with respect to gravity effects so the influence of the magnetic fields to the geometry will be negligible. 
In this framework the equations are considerably simplified and the problem
consists in finding the solutions for non-minimally coupled magnetic fields on  a static spherically symmetric spacetime. The static spherically symmetric spacetime is described by the Schwarzschild metric
\begin{equation}
 ds^{2}=-\Big(1-\frac{2M}{r}   \Big)   dt^{2}+ \frac{dr^{2}}{\Big(1-\frac{2M}{r}   \Big)}+ r^{2}(d\theta^{2} +  \sin^{2} \theta d\phi^{2}),
 \label{metrika}
\end{equation}
in this coordinates the Maxwell tensor becomes
\begin{equation}
 F_{\mu\nu}=\begin{pmatrix}
    0       &    0    &   0   &    0 \\
    0       &    0    & \frac{rB_{\phi}}{\sqrt{1-\frac{2M}{r}}} &    -\frac{r\sin \theta B_{\theta}}{\sqrt{1-\frac{2M}{r}}} \\
    0       &    -\frac{rB_{\phi}}{\sqrt{1-\frac{2M}{r}}}    &    0      &    r^{2}\sin \theta B_{r} \\
    0       &   \frac{r\sin \theta B_{\theta}}{\sqrt{1-\frac{2M}{r}}}     &    -r^{2}\sin \theta B_{r}   &    0
\end{pmatrix},
\end{equation}
now we can solve the equation \eqref{sacuvanje}. The Schwarzschild metric is a vacuum solution and it follows that $R=0$ and $R^{\mu \nu}=0$, the only remaining part of $\mathcal{R^{\mu\nu\rho\sigma}}$
is the Riemann tensor:
\begin{equation}
 \mathcal{ R}^{\mu\nu\rho\sigma}=q_{3}R^{\mu\nu\rho\sigma},
\end{equation}
and equation \eqref{sacuvanje} becomes
\begin{equation}
 \nabla_{\mu}F^{\mu\nu} + q_{3}\nabla_{\mu}(R^{\mu\nu\rho\sigma}F_{\rho\sigma})=0.
\end{equation}
In order to inspect the magnetic fields in the vicinity of the Schwarzschild metric it is convenient to consider a realistic configuration of magnetic fields which is of physical interest. Since the Schwarzschild metric is asymptotically flat, at distances which are considerably larger than the Schwarzchild horizon the black hole spacetime will approach to the Minkowski spacetime and the solution of \eqref{sacuvanje} will be asymptotically given by the problem of magnetic fields on the flat spacetime which is the same as in the minimally coupled case. Due to its relevance 
as the model for the galactic magnetic field distribution, we will focus on the field configuration which leads to magnetic dipole on the flat spacetime. 

In this way we can further simplify the problem and set
$B_{\phi}=0$, $B_{r}(r,\theta,\phi)=B_{r}(r,\theta)$, $B_{\theta}(r,\theta,\phi)=B_{\theta}(r,\theta)$ and 
$B_{\theta}(r,\theta)=\tan \theta B_{r}(r,\theta)/2$.
By putting all this assumptions in the equation \eqref{sacuvanje} we get 
\begin{multline}
 r(r-2M)(r^{3}-2Mq_{3})\frac{dB_{r}(r,\theta)}{dr}\\
 -(10M^{2}q_{3}-r^{4}+M(r^{3}-4q_{3}r))B_{r}(r,\theta) \\
 = \frac{2}{\tan \theta} \sqrt{1-\frac{2M}{r}}r(4Mq_{3} + r^{3})\frac{dB_{r}(r,\theta)}{d\theta}.
 \label{jednadzba}
\end{multline}
The equation \eqref{jednadzba} can be separated by using $B_{r}(r,\theta)=B_{Rad}(r)\Theta(\theta)$ and by doing so we get two equations:
\begin{equation}
\frac{d\Theta(\theta)}{d \theta}-C\tan \theta \Theta(\theta)=0,
\label{theta}
\end{equation}
\begin{equation}
 \frac{dB_{Rad}(r)}{dr} - \frac{  A_{1}(r) + CA_{2}(r)}{r(r-2M)(r^{3}-2Mq_{3})} B_{Rad}(r)=0,
 \label{radijalna}
\end{equation}
where
\begin{equation}
 A_{1}(r)=\Big( 10M^{2}q_{3} - r^{4} + M(r^{3}-4q_{3}r)\Big),
\end{equation}
\begin{equation}
 A_{2}(r)=2\sqrt{ 1-\frac{2M}{r}}r(4Mq_{3}+r^{3}),
\end{equation}
and $C$ is the separation constant.
The equation \eqref{theta} can be easily solved and the solution is:
\begin{equation}
 \Theta(\theta)=\Theta_{0} \cos^{-C} \theta,
 \label{thetajedn}
\end{equation}
where $\Theta_{0}$ is the integration constant.
When we are far away from the black hole horizon, $2M \ll r$, the equation \eqref{thetajedn} remains the same and the dipole magnetic field configuration requires the choice $C=-1$, while \eqref{radijalna} simplifies leading to a well known dipole solution: $B_{RAD}=constant/r^{3}$. The considered system thus reduces to the case of magnetic dipole on a flat spacetime in the asymptotically flat limit. In the next section we will consider the numerical solutions of 
\eqref{radijalna} around the black hole horizon. 

\section{Results}

\begin{figure}[h]
    \includegraphics[scale=0.50]{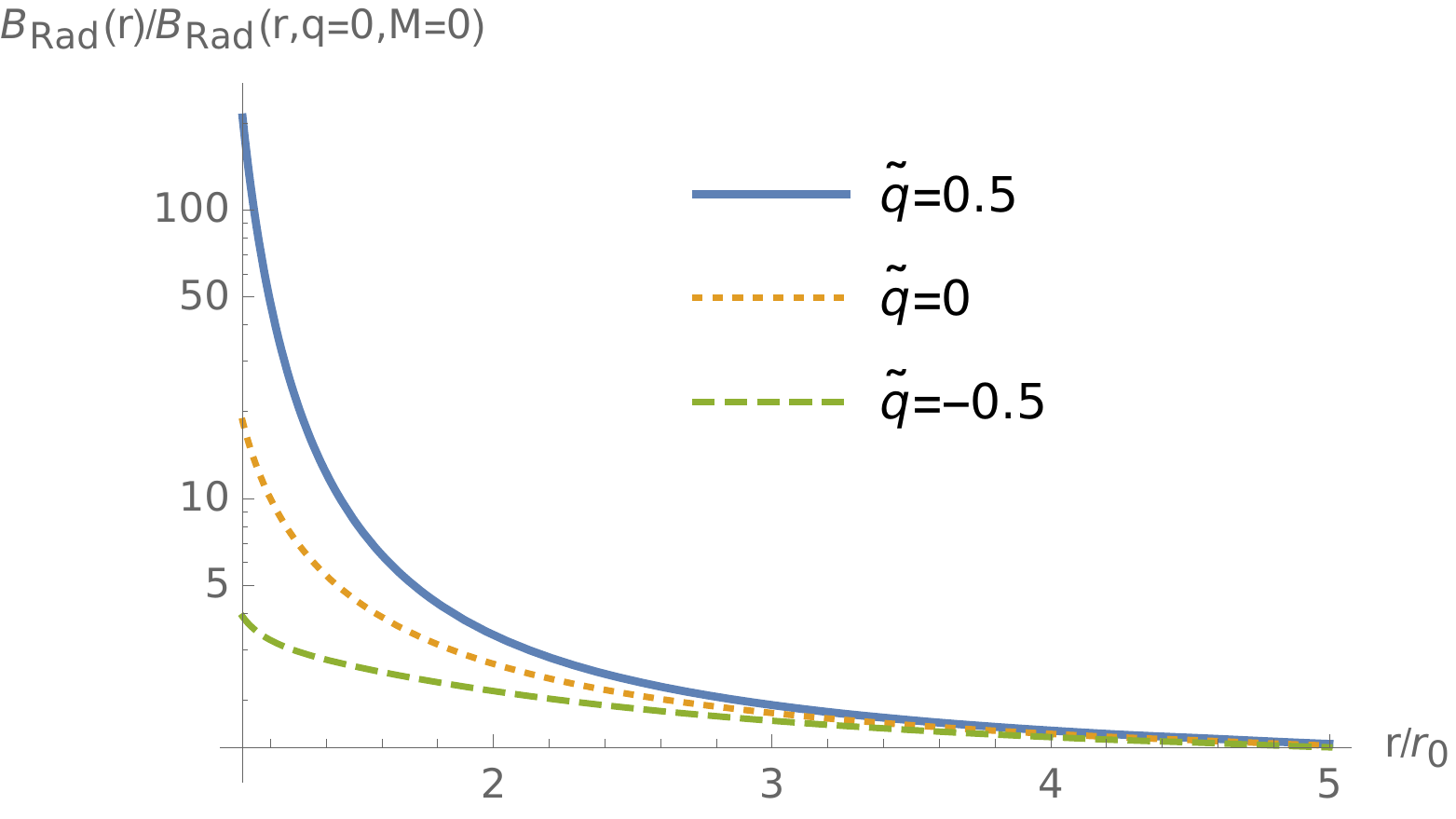}
       \caption{The radial component of
       the magnetic field $B_{r}(r)$ scaled with
       respect to the magnetic dipole field as a
       function of
       $r/r_{0}$ for different values of $\tilde{q}$.
       The full
       blue line represent the $\tilde{q}=0.5$, the
       orange dotted line is the GR (minimal coupling)
       case, while the green
       dashed line is the negative case with
       $\tilde{q}=-0.5$.
       For positive values of $\tilde{q}$ the 
       magnetic field experience a significant
       amplification near the horizon, on the other
       hand for negative $\tilde{q}$ the
       field slowly diminishes approaching the horizon.
       It is also visible that by going further away
       from the horizon the fields soon approach to
       the GR (minimal coupling) case.
       } \label{Bfunr}
\end{figure}

Firstly, we will rescale the equation \eqref{radijalna} to work in dimensionless variables
\begin{equation}
 r\rightarrow \tilde{r}=\frac{r}{r_{0}}, \qquad M \rightarrow \tilde{M}=\frac{M}{r_{0}}, \qquad q_{3}\rightarrow \tilde{q}=\frac{q_{3}}{r_{0}^{2}},
 \label{rescale}
\end{equation}
where $r_{0}$ is the free scaling parameter and in our case we set $r_{0}=2M$ - the Schwarzschild radius. In order to numerically  solve the equation \eqref{radijalna} we need one boundary condition for
which we choose $B_{Rad}(r=100r_{0})=10^{-10}$T, where
this value was motivated by the typical value of the
magnetic field in our galaxy, which we assume to be
approached on distances much larger than $r_{0}$
\cite{magnetic}. As the obtained solutions will be
scaled with respect to the magnetic dipole on flat
spacetime with the same boundary condition, this
numerical assumption will not influence our results.
The solution to the equation \eqref{radijalna} for
different parameters is depicted in FIG. \ref{Bfunr}.
\begin{figure}[h]
    \includegraphics[scale=0.50]{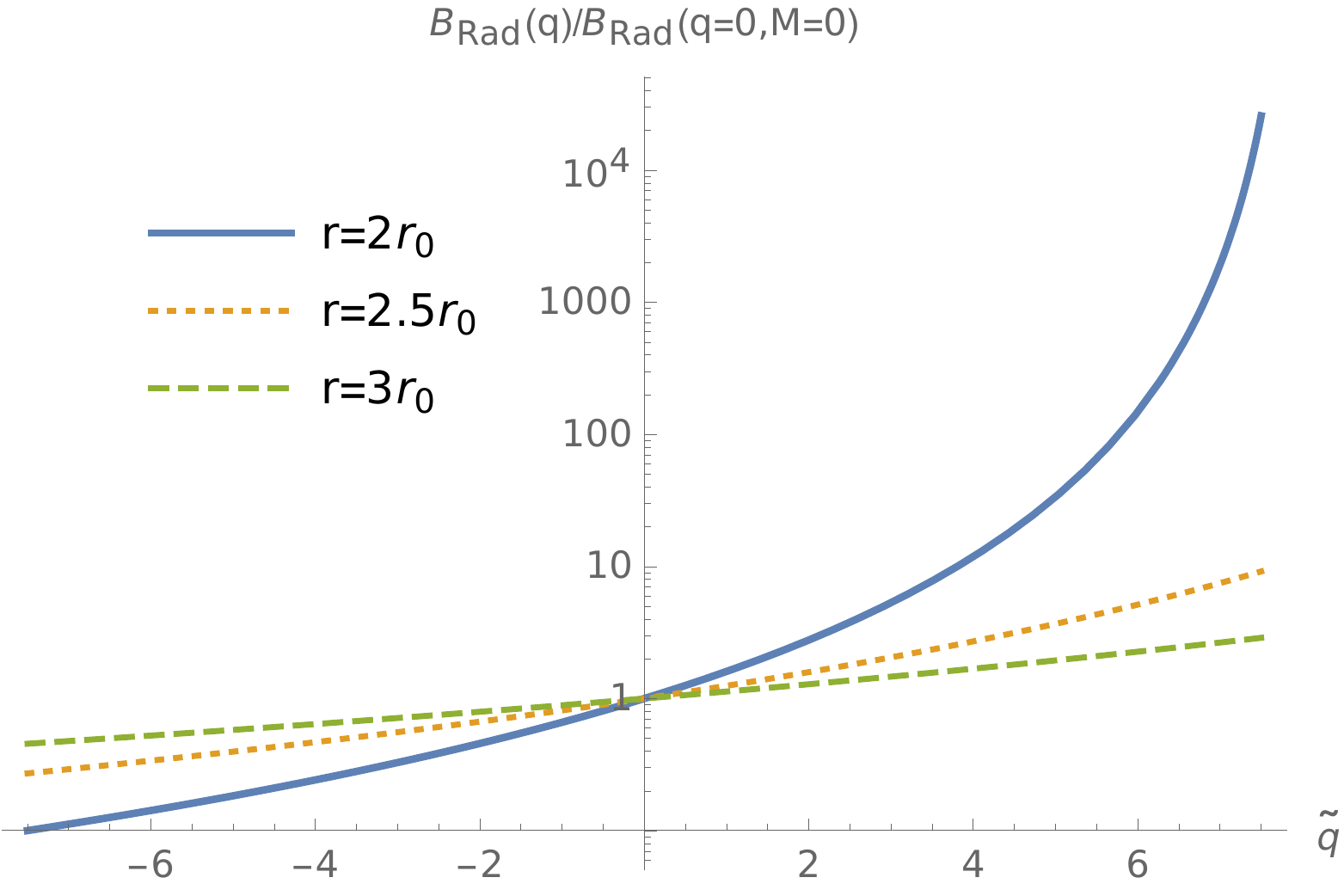}
       \caption{The radial component of
       the magnetic field $B_{r}(q)$ scaled with
       respect to the magnetic dipole field as a
       function of
       $\tilde{q}$ for fixed radial value. 
       The full blue line represent the magnetic field
       as a function of $\tilde{q}$ for the fixed
       radial value $r=2r_{0}$, the
       orange dotted line is for the fixed radial 
       value $r=2.5r_{0}$, while the green
       dashed line is the $r=3r_{0}$ case. Here again
       it is visible that near the horizon the
       magnetic fields experience a high 
       amplification, by going further away from the
       horizon the impact of $\tilde{q}$ becomes 
       suppressed. By increasing $\tilde{q}$ the
       magnetic field blows up at the effective source
       singularity $r=(2Mq_{3})^{1/3}$.} 
       \label{Bfunq}
\end{figure}
It is also interesting to plot the magnetic field as
a function of $\tilde{q}$ for a fixed radial value where it
can be shown the relative change of magnitude by
increasing or decreasing the parameter $\tilde{q}$, as shown in FIG. \ref{Bfunq}. 
In the vicinity of the Schwarzschild radius the relative
change of magnitude of the magnetic field is
drastically increased by increasing $\tilde{q}$, but further
away from the Schwarzschild radius this dependence
starts to be less pronounced. This means that
at sufficiently large $r$ the nonminimal coupling
effect of gravity and electromagnetic approximately disappears.
It can be observed that equation \eqref{radijalna}
apart from two known singularities -- the coordinate
singularity at the horizon $r=2M$, and the physical
singularity at $r=0$ -- leads to an additional
singularity at $r=(2Mq_{3})^{1/3}$. It is important
to emphasize that the singularity at $r=0$ is no
longer only a spacetime singularity, but is now also a
field source singularity. The geometrical nature of
this singularity can be shown by using the standard 
analysis of divergence of Kretschmann scalar,
$R^{\mu \nu \rho \sigma}R_{\mu \nu \rho \sigma}$, at
$r=0$. The additional existence of field source singularity at the same spacetime point can be simply observed by taking the limit of
Minkowski spacetime where the magnetic dipole
solution, $B_{RAD}=constant/r^{3}$, is singular at $r=0$. On the other
hand, the new singularity at $r=(2Mq_{3})^{1/3}$ is 
obviously not related to the curvature singularity
on the spacetime described by equation
\eqref{metrika}, but is a consequence of introducing
the vacuum polarization. This can be understood 
as a result of the additional terms which enter in 
the electromagnetic field equation due to the vacuum
polarization effect. These additional terms can be 
viewed as a new effective field configuration
by defining $F^{\mu \nu}_{eff}\equiv F^{\mu \nu}
+\mathcal{R}^{\mu\nu\rho\sigma}F_{\rho\sigma}$. 
In this sense the new singularity can be seen as
source singularity of the effective quantum 
corrected electromagnetic field distribution.
We note that under the assumptions taken in this
work this singularity will always be hidden within
the black hole event horizon. This follows from
the fact that under the
approximation of negligible feedback on the
spacetime, $|q_{3}|\ll M^{2}$, the condition
$(2|q_{3}|M)^{1/3}\ll 2M$ needs to be satisfied.
The presence of such singularity could become 
important in the case of strong electromagnetic 
fields where the interplay between the field 
 distribution and the spacetime geometry could no
 longer be neglected. If this singularity would
 still persist in the fully self-consistent 
 analysis then it would be necessary to go beyond
 one loop correction used in equation
 \eqref{oneloop}. 
 
 \section{Implication for physical systems}
 
 As can be seen by the results presented in the
 previous section, visible in FIG. \ref{Bfunr}, 
 the positive sign of the vacuum polarization 
 parameter $\tilde{q}$ typically leads to 
 amplification of dipole magnetic fields near the black
 hole horizon with respect to both, magnetic dipole
 solution and minimally coupled ($\tilde{q}=0$) 
 magnetic field on Schwarzschild spacetime. In the
 opposite case with negative $\tilde{q}$ magnetic 
 fields are suppressed. However, these effects 
 become negligible soon as we move away from the 
 horizon. From FIG. \ref{Bfunq} we conclude that a
 relevant amplification of magnetic fields
 requires the coupling coefficient to be at least of
 the order of magnitude, $\tilde{q} \approx 1$. 
 From definition \eqref{rescale} we see that this 
 implies $q_{3}\approx r_{0}^2$, which is a huge 
 value of a coupling parameter if the usual 
 astrophysical values of $r_{0}$ are assumed. It can be seen from FIG. \ref{Bfunq} that for $\tilde{q}$ of the same order of magnitude, the magnetic fields can be
 enhanced for more than four orders of magnitude, therefore in principle leading to practically observable
 consequences.
 On the other hand, in most
 systems apart from black holes the gravitational
 curvature effects will be negligible and thus would 
 also be the effects of vacuum polarization related
 to this parameter. For instance in the case of 
 the Earth with $\tilde{q}=1$ where $R_{Earth}/r_{0}
 \approx 10^{9}$ and the Sun $R_{Sun}/r_{0}
 \approx 10^{5}$ the resulting ratios
 $B_{Rad}(R_{Earth})/B_{Rad}(R_{Earth},\tilde{q}=0,M=0)$ and
 $B_{Rad}(R_{Sun})/B_{Rad}(R_{Sun},\tilde{q}=0,M=0)$
 are practically indistinguishable from the GR case with
 $\tilde{q}=0$. The difference is more pronounced
 in the case of neutron stars where $R_{NS}/r_{0}
 \approx 17$ so the resulting variation from the
 $\tilde{q}=0$ is around $0.1\%$. Therefore, even the high values of the coupling parameter, making it of the order of macroscopic lengths -- as given by the square of Schwarzschild radius -- would lead 
 to practically detectable modifications of magnetic fields only in the vicinity of black holes. The strategy to potentially observe the electromagnetic effects of vacuum polarization in a gravitational field -- or at least to put constraints on the coupling parameter $q_{3}$ -- would be therefore to systematically search for the variations of magnetic field around  the horizons of black holes. 
\\ \\
Although the value of the coupling $q_{3}$ should be considered as a free parameter for the systems studied in this work -- involving much more complicated configurations than a single photon placed in a gravitational field, we know \cite{drummond} that its  value needs to be at least of the order of the square of the Compton wavelength for electron, $\lambda_{c}$. This most conservative option should thus be considered with a specific attention. In this case, $\tilde{q}=(\lambda_{c}/
r_{0})^{2}$. For all astrophysical systems then clearly follows $\tilde{q}
\ll 1$ and therefore no observable effect could be expected. However, it has been speculated \cite{primordial1, primordial2} -- and this option has gained a lot of popularity recently -- that black holes of microscopic sizes could be created in the conditions of the early Universe. Since the minimal Schwarzschild radius of primordial black holes is expected to be larger only from the Planck length, if primordial black holes exist one would also expect that at least for some of them $r_{0} \leq \lambda_{c}$ and their existence would lead to modifications of primordial magnetic fields which could not be neglected. In fact, assuming the existence of primordial black holes comparable to Planck length would lead to $\tilde{q} \approx 10^{46}$ which would have a tremendous effect on the strengths and distribution of primordial magnetic fields, but such strong coupling regimes are much beyond the weak field and one-loop approximations used in this work. In any case, the investigations 
assuming the existence of such primordial black holes and discussing their distribution should  necessary
consider the effects on the primordial magnetic fields due to the vacuum polarization. Leaving this regime for now aside, let us briefly discuss the potential consequences of primordial black holes for which $r_{0} \sim \lambda_{c}$. From the results presented in the previous section it follows that such 
primordial black holes 
would lead to an effective amplification or suppression of primordial dipole magnetic field in the close vicinity of their microscopic Schwarzschild radius potentially for many orders of magnitude.
The primordial black holes would subsequently evaporate as the Universe evolves, but the change in the local field strengths in the vicinity of regions where they were previously existing, would in principle still be present even today. They would be visible as  strongly localized significant departures from the average magnetic field strengths. Searching and analyzing such field patterns in the observed astrophysical magnetic fields could be used for testing the discussed vacuum polarization effects and setting the constraints on its characteristic parameters, but also the existence and distribution of primordial black holes. Taking into account that the question of magnetogenesis and the
subsequent evolution of cosmological magnetic fields represents an interesting and important
problem in cosmology \cite{magneto1, magneto2, magneto3, natty} these considerations are also important since they could significantly  influence it. 
\section{Conclusion}
Motivated by the effect of vacuum polarization for
photons in gravitational field, which leads to the
non-minimal coupling between gravity and
electromagnetism, we studied its potential
consequences on magnetic fields present in the
Universe. By assuming that magnetic fields are 
sufficiently weak so that their back-reaction on
spacetime can be ignored, which is the case for 
characteristic strengths of astrophysical magnetic
fields, we solve the resulting field equation
on the Schwarzschild spacetime. The obtained solutions
demonstrate that in the vicinity of the Schwarzschild
horizon the magnetic fields can become significantly
amplified or damped - with respect to flat spacetime
and minimal coupling case - depending on the value 
and sign of the characteristic coupling constant. 
Furthermore, by moving away from the Schwarzschild
horizon the obtained magnetic fields rapidly approach
to the magnetic dipole configuration on flat 
spacetime. We have also discussed the singularities 
present in the considered system - where apart from
the well known curvature and field source
singularities at $r=0$ a new singularity at 
$r=(2Mq_{3})^{1/3}$ arises. We have argued that this
singularity can be understood as the result of the
new effective quantum corrected electromagnetic field
distribution $F^{\mu \nu}_{eff}\equiv F^{\mu \nu}
+\mathcal{R}^{\mu\nu\rho\sigma}F_{\rho\sigma}$. 
Moreover, under the assumption of negligible 
feedback of the magnetic fields on the spacetime,
this singularity will always be hidden inside the 
event horizon. Finally, we have discussed how such
effects can be observable in principle by
systematically considering the variations of magnetic 
fields around the horizons of black holes. Also,
 this effect could be strongly pronounced in the 
 context of cosmological magnetic fields where the 
 existence of primordial black holes could lead
 to strong modifications of magnetic field
 configurations and strengths. These effects would
 manifest, depending on the sign of the coupling parameter, either as  pronounced peaks in the distribution of astrophysical magnetic fields or in 
 the opposite case as localized regions of negligible
 magnetic field strengths significantly departing from 
 the avarage values. 
 Therefore, the 
 considered effects can have important consequences
 on the questions of magnetogenesis and evolution of
 magnetic fields in the early Universe.

\end{document}